\documentclass[aps,prl,reprint,superscriptaddress,longbibliography]{revtex4-2}
\usepackage{graphicx}
\usepackage{color}
\usepackage[tight]{units}
\usepackage{epstopdf}
\usepackage[normalem]{ulem}
\DeclareGraphicsExtensions{.pdf,.jpg,.png,.eps}
\usepackage{amsfonts}
\usepackage{amsmath}
\usepackage{amssymb}
\usepackage{color}
\usepackage{color}
\usepackage{lineno}
\usepackage{epstopdf}
\usepackage{adjustbox}

\usepackage[T1]{fontenc}
\usepackage{hyperref}
\usepackage{threeparttable}

\newcommand{\xs}

\begin{document}

\title{\textrm{From Chemical Complexity to Tunable Magnetic Ordering in Highly Disordered High-Entropy Spinel Oxides}}

\author{Neha Sharma}
\affiliation{Department of Physics and Material Science, Thapar Institute of Engineering and Technology, Patiala 147004, India}

\author{Sushanta Mandal}
\author{Nikita Sharma}
\affiliation{Department of Physics and Material Science, Thapar Institute of Engineering and Technology, Patiala 147004, India}
\author{Amritpal}
\affiliation{UGC-DAE Consortium for Scientific Research, Khandwa Road, Indore 452001, Madhya Pradesh, India}
\author{Sangeeta Thakur}
\affiliation{Institut für Experimentalphysik, Freie Universität Berlin, Arnimallee 14, 14195  Berlin, Germany}
\author{Viktor Ukleev}
 \author{Chen Luo}
\author{Florin Radu}
\affiliation{Helmholtz Zentrum Berlin für Materialien und Energie, Albert-Einstein Straße 15, 12489 Berlin, Germany}

\author{S. D. Kaushik}
\affiliation{UGC-DAE Consortium for Scientific Research, Mumbai Centre, 246-C CFB, BARC Campus, Trombay, Mumbai-400085, India}
\author{Tirthankar Chakraborty}
\affiliation{Department of Physics and Material Science, Thapar Institute of Engineering and Technology, Patiala 147004, India}
\author{Sanjoy Kr. Mahatha}
\affiliation{UGC-DAE Consortium for Scientific Research, Khandwa Road, Indore 452001, Madhya Pradesh, India}
\author{Denis Pelloquin}
\affiliation{Laboratory Crismat, UMR6508 CNRS, Normandie University, ENSICAEN, UNICAEN, 6 bd Maréchal Juin, 14050 Caen cedex 4, France}
\author{Sourav Marik}
\email[]{soumarik@thapar.edu}
\affiliation{Department of Physics and Material Science, Thapar Institute of Engineering and Technology, Patiala 147004, India}


\begin{abstract}

High-entropy stabilization chemistry is redefining materials design by transforming configurational disorder, arising from the deliberate incorporation of multiple principal cations, into a thermodynamic advantage that promotes phase stability and enables emergent functionalities. In this work, we investigate the evolution of magnetic ordering in spinel-type high entropy oxides by systematically varying the cation composition of the B site within a fixed high-entropy A-site matrix, (Ni$_{0.2}$Mg$_{0.2}$Co$_{0.2}$Cu$_{0.2}$Zn$_{0.2}$)B$_2$O$_4$. Upon introducing multicomponent B-site configurations, we uncover a strikingly linear dependence of the magnetic transition temperature (T$_C$) on the T$_C$s of the corresponding single B-site high-entropy systems. Remarkably, this trend persists even in highly complex (Ni$_{0.2}$Mg$_{0.2}$Co$_{0.2}$Cu$_{0.2}$Zn$_{0.2}$)(Cr$_{0.2}$Mn$_{0.2}$Fe$_{0.2}$Ga$_{0.2}$X$_{0.2}$)$_2$O$_4$, X = Al and Ti. Despite the material's extremely high degree of disorder, absence of a dominant magnetic ion or a straightforward superexchange pathway, detailed magnetization measurements, low-temperature X-ray magnetic circular dichroism, and neutron powder diffraction studies reveal robust long-range ferrimagnetic ordering. These results reveal an emergent predictability in ferrimagnetic high-entropy spinel oxides, where, despite extreme configurational disorder and competing interactions, robust ferrimagnetic order can arise from, rather than be
hindered by, extreme configurational disorder. This establishes a pathway for predictively tuning magnetic transition temperatures in high-entropy oxides beyond conventional ordered systems.
\end{abstract}

\maketitle

\section{Introduction}
In conventional solid-state chemistry, structural order has long been equated with stability and functionality \cite{1}. Crystalline symmetry, well-defined cation sublattices, and predictable phase diagrams have historically underpinned the design of magnetic materials. Studies on conventionally doped systems have shown that disorder can disrupt long-range magnetic order, often leading to glassy magnetic states, compositional inhomogeneities, or the emergence of Griffiths-like phases \cite{2, 3}. However, an intriguing question arises: what if we intentionally forgo structural order, not to introduce randomness, but to engineer purposeful complexity? This question has gained prominence with the advent of entropy-stabilized materials, where configurational disorder is no longer treated as a defect but rather as a stabilizing thermodynamic force. High-entropy materials, which typically incorporate five or more principal elements into a single crystallographic sublattice, push the boundaries of traditional solid solutions \cite{4, 5, 6, 7, 8}. The thermodynamic stabilization of such disordered yet crystalline materials is governed by the Gibbs free energy equation:
\[\Delta G_{\text{mix}} = \Delta H_{\text{mix}} - T \Delta S_{\text{mix}}\]

T = temperature and $\Delta H_{\text{mix}}$ = mixing enthalpy, $\Delta S_{\text{mix}}$ = mixing entropy and $\Delta G_{\text{mix}}$ = Gibbs free energy. In high-entropy materials, the large positive entropy contribution can dominate the energetics, promoting the formation of single-phase solid solutions with simple crystal structures despite their high compositional complexity. For a compound with chemical composition $A_{\alpha}B_{\beta}O_{\gamma}$, the configurational entropy $S_{\text{mix}}$ is expressed as \cite{9} 
\begin{align}
S_{\text{mix}} = -R \Bigg[ \alpha \sum_{i=1}^{N} a_i \ln a_i \notag 
 + \beta \sum_{j=1}^{M} b_j \ln b_j \notag 
& + \gamma \sum_{k=1}^{P} c_k \ln c_k 
\Bigg]
\label{eq:Sconfig}
\end{align}
Here, $a_i$, $b_j$, and $c_k$ = mole fractions of elements occupying the A-site, B-site, and O-site, respectively, and $\alpha$, $\beta$, $\gamma$ are the corresponding stoichiometric coefficients in the chemical formula.

\begin{figure*}
    \centering
    \includegraphics[width=1.00\linewidth]{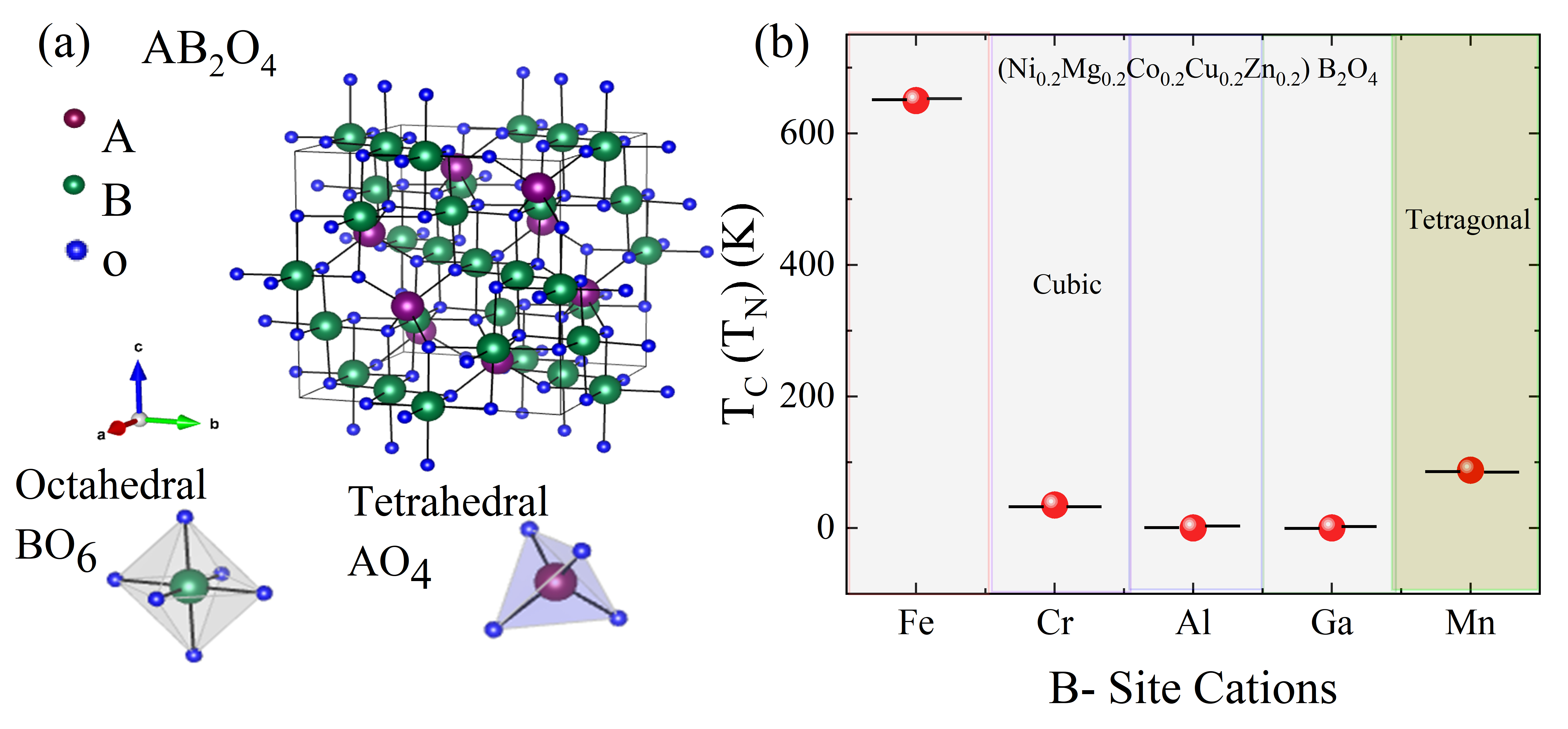}
    \caption{(a) Crystal structure of cubic spinel AB${_2}$O${_4}$ structure. (b) Magnetic transition temperatures and crystal structures for (Ni$_{0.2}$Mg$_{0.2}$Co$_{0.2}$Cu$_{0.2}$Zn$_{0.2}$)B$_2$O$_4$ materials, where B = Cr, Fe, Al, Ga, and Mn. Magnetic transition temperatures for single B-site compositions are taken from ref \cite{10, 11, 12}}
    \label{fig1}
\end{figure*}

\begin{figure*}
    \centering
    \includegraphics[width=1.00\linewidth]{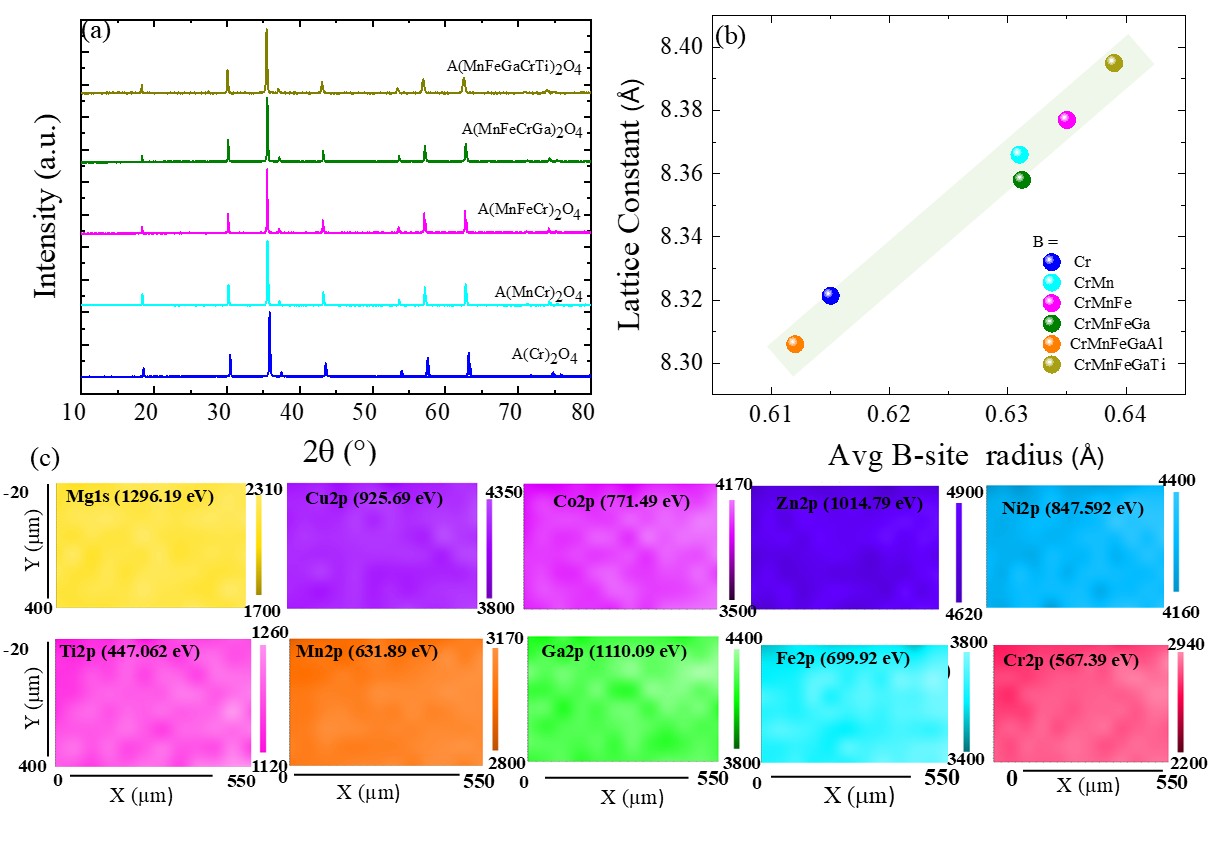}
    \caption{(a) Room temperature X-ray diffraction patterns for (Ni$_{0.2}$Mg$_{0.2}$Co$_{0.2}$Cu$_{0.2}$Zn$_{0.2}$)B$_2$O$_4$ materials, where B = Cr, Mn$_{0.5}$Cr$_{0.5}$, Mn$_{0.33}$Cr$_{0.33}$Fe$_{0.33}$, Mn$_{0.25}$Cr$_{0.25}$Fe$_{0.25}$Ga$_{0.25}$, and Mn$_{0.2}$Cr$_{0.2}$Fe$_{0.2}$Ga$_{0.2}$Ti$_{0.2}$. (b) Variation of lattice constant (a = b = c) with average B-site cationic radius. A monotonic increase in the lattice constant is observed with increasing average B-site ionic radius, consistent with Vegard's law. (c) Room-temperature X-ray photoelectron spectroscopy (XPS) mapping for compositionally most complex sample, (Ni$_{0.2}$Mg$_{0.2}$Co$_{0.2}$Cu$_{0.2}$Zn$_{0.2}$)(Cr$_{0.2}$Mn$_{0.2}$Fe$_{0.2}$Ga$_{0.2}$Ti$_{0.2}$)$_2$ O$_4$, performed with a 50 $\mu$m spot size, reveals a uniform spatial distribution of all constituent elements.}
    \label{fig1}
\end{figure*}

Among high-entropy materials, high-entropy oxides (HEOs) have emerged as a powerful platform for the exploration of structural, electronic, and magnetic phenomena, where chemical complexity can lead to emergent and often counterintuitive properties \cite{13, 14, 15, 16, 17, 18, 19, 20, 21, 22, 23, 24}. Since the initial demonstration in rocksalt-type HEOs \cite{25}, this concept has expanded to encompass a range of structural motifs, including perovskite \cite{26}, fluorite \cite{27}, and spinel lattices \cite{28}. Concurrently, an extensive array of functionalities has been explored, covering mechanical \cite{29}, thermal \cite{30, 31}, catalytic \cite{32}, optical \cite{33}, and magnetic \cite{14} properties. In this study, we present a design strategy to modulate the magnetic properties in spinel-type high-entropy oxides. The spinel materials (AB$_2$O$_4$, Fig. 1(a)) are particularly attractive due to their wide range of applications. Recent investigations of spinel HEOs incorporating 3d transition metals have reported ferri- or antiferromagnetic transitions, with critical temperatures highly sensitive to the specific cationic makeup of the A and B sites (Fig. 1(b)) \cite{10, 16, 28}. Yet, a central challenge remains: can we systematically and predictively tune such magnetic transitions via compositional design? In this study, we propose a design framework based on spinel AB$_2$O$_4$ structure with a fixed high-entropy A-site matrix, [Ni$_{0.2}$Mg$_{0.2}$Co$_{0.2}$Cu$_{0.2}$Zn$_{0.2}$], while systematically varying the B-site cations among Cr, Mn, Fe, Ga, Al, and Ti. Each parent compound exhibits a distinct magnetic transition temperature reflecting the intrinsic magnetic character of the B-site ion. Strikingly, upon forming multicomponent B-site compositions, we observe that the resulting transition temperatures follow a nearly linear, arithmetic dependence on those of the parent systems. This simple scaling persists even in highly complex compositions such as (Ni$_{0.2}$Mg$_{0.2}$Co$_{0.2}$Cu$_{0.2}$Zn$_{0.2}$)(Cr$_{0.2}$Mn$_{0.2}$Fe$_{0.2}$Ga$_{0.2}$Ti$_{0.2}$)$_2$O$_4$, which exhibits robust long-range magnetic ordering despite the absence of a dominant magnetic ion or a well-defined exchange pathway. Our results establish an intriguing example of predictable magnetic behavior arising from extreme chemical disorder, providing a new route to rationally design functional oxide materials through high-entropy compositional engineering.

\section{Results and Discussion}
Figure 2(a) presents the room-temperature X-ray diffraction (RT-XRD) patterns for the spinel-type high-entropy oxides with nominal compositions of (Ni$_{0.2}$Mg$_{0.2}$Co$_{0.2}$Cu$_{0.2}$Zn$_{0.2}$)B$_2$O$_4$, where B = Cr, Mn$_{0.5}$Cr$_{0.5}$, Mn$_{0.33}$Cr$_{0.33}$Fe$_{0.33}$, Mn$_{0.25}$Cr$_{0.25}$Fe$_{0.25}$Ga$_{0.25}$, and Mn$_{0.2}$Cr$_{0.2}$Fe$_{0.2}$Ga$_{0.2}$Ti$_{0.2}$. The XRD patterns for single B-site compositions (B = Fe, Ga, Al, Mn) are provided in the supporting information (SI, Figure S1) file. All diffraction patterns confirm the formation of phase-pure materials. Rietveld refinement of the RT-XRD reveals that all compositions crystallize in the cubic spinel structure with space group $Fd\overline{3}m$. Figure 2(b) illustrates the evolution of the lattice parameter as a function of the average ionic radius at the B-site. A monotonic increase in the lattice constant is observed with increasing average B-site ionic radius, consistent with Vegard's law.

\begin{figure*}
    \centering
    \includegraphics[width=1.00\linewidth]{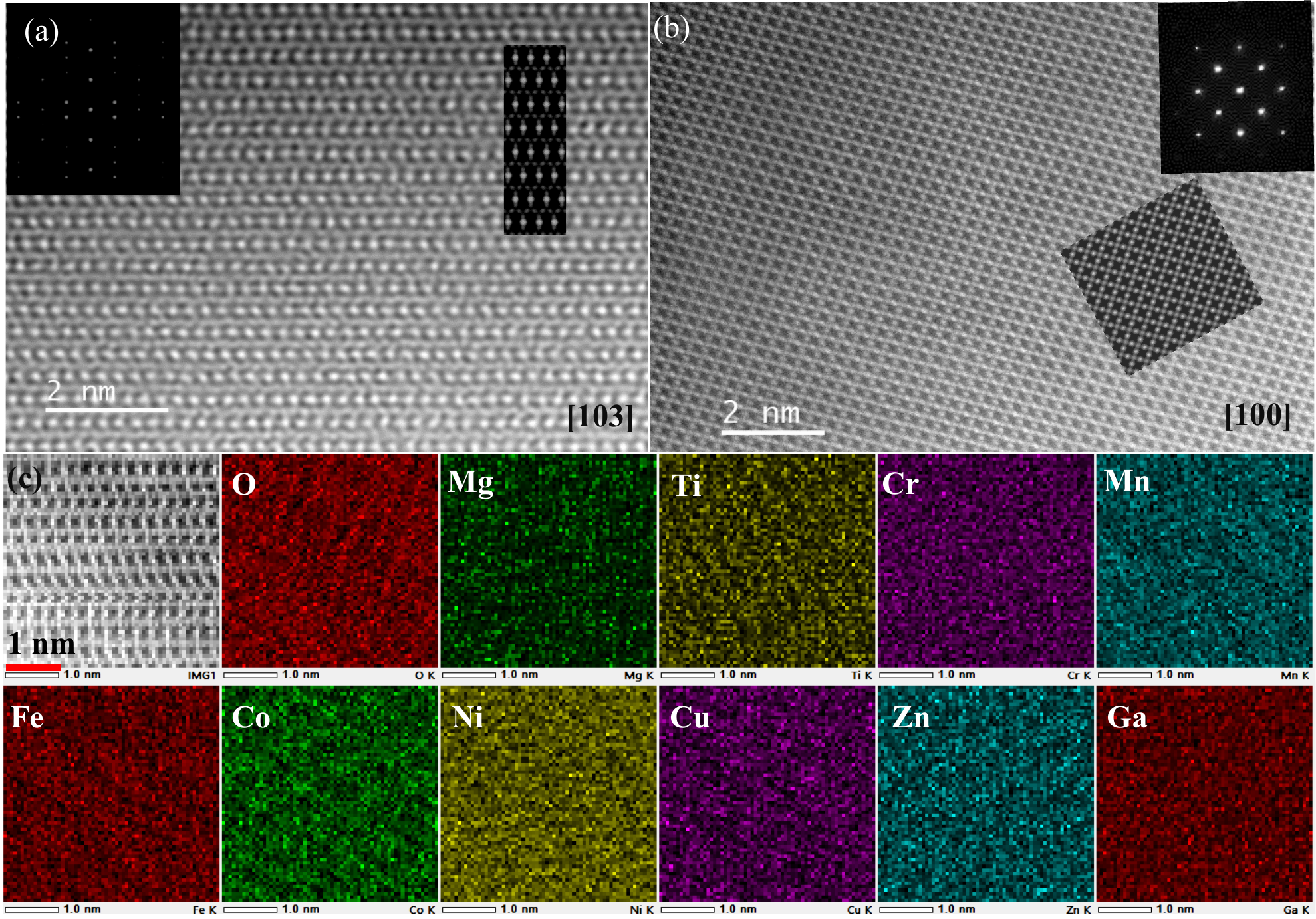}
    \caption{Room-temperature high-angle annular
dark-field Scanning Transmission Electron Microscopy (HAADF-STEM) image acquired along the (a) [103] and (b) [100] zone axis for compositionally most complex sample, (Ni$_{0.2}$Mg$_{0.2}$Co$_{0.2}$Cu$_{0.2}$Zn$_{0.2}$)(Cr$_{0.2}$Mn$_{0.2}$Fe$_{0.2}$Ga$_{0.2}$Ti$_{0.2}$)$_2$ O$_4$. The corresponding fast Fourier transform (FFT) pattern exhibits distinct reflections consistent with cubic symmetry. Simulated HAADF-STEM images, shown alongside the experimental data, further support the structural feature. The periodic lattice arrangements are in excellent agreement with the expected cation distribution in a spinel lattice, indicating that the high configurational disorder does not disrupt the long-range crystallographic order. (c) Atomic-scale EDS elemental mapping for the same sample, highlighting the excellent chemical homogeneity on the nanoscale.}
    \label{fig3}
\end{figure*}

\begin{figure*}
    \centering
    \includegraphics[width=1.00\linewidth]{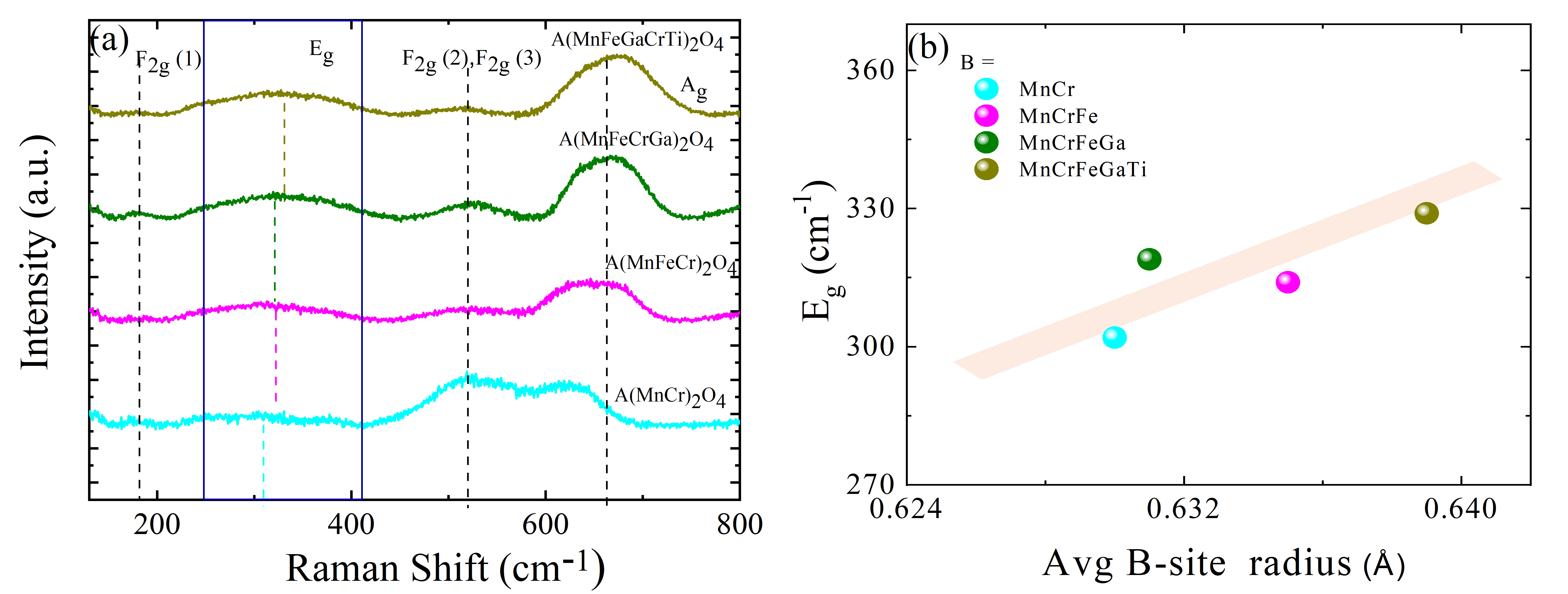}
    \caption{Room temperature Raman spectra for (Ni$_{0.2}$Mg$_{0.2}$Co$_{0.2}$Cu$_{0.2}$Zn$_{0.2}$)B$_2$O$_4$ materials, where B = Mn$_{0.5}$Cr$_{0.5}$, Mn$_{0.5}$Fe$_{0.5}$, Mn$_{0.33}$Cr$_{0.33}$Fe$_{0.33}$, Mn$_{0.25}$Cr$_{0.25}$Fe$_{0.25}$Ga$_{0.25}$, and Mn$_{0.2}$Cr$_{0.2}$Fe$_{0.2}$Ga$_{0.2}$Ti$_{0.2}$. In agreement with the cubic spinel oxides, all the samples exhibit five Raman-active modes. (b) Variation in octahedra-linked E$_g$ modes with the average radius of the B-site cations.}
    \label{fig4}
\end{figure*}

To further assess the microstructure, chemical homogeneity, and elemental distribution, field-emission scanning electron microscopy (FESEM) combined with energy-dispersive X-ray spectroscopy (EDS) mapping was performed for all compositions and is shown in the SI (Fig. S2). The EDS mappings confirm a homogeneous distribution of all constituent elements at the micrometer scale, indicative of excellent chemical mixing. X-ray photoelectron spectroscopy (XPS) mapping, performed on compositionally most complex sample, (Ni$_{0.2}$Mg$_{0.2}$Co$_{0.2}$Cu$_{0.2}$Zn$_{0.2}$)(Cr$_{0.2}$Mn$_{0.2}$Fe$_{0.2}$Ga$_{0.2}$Ti$_{0.2}$)$_2$ O$_4$ using a 50 $\mu$m spot size beam (Fig. 2 (c)) reveals a uniform spatial distribution of all constituent elements. In agreement with energy-dispersive X-ray spectroscopy (EDS) analysis, the XPS mappings confirm a high degree of chemical homogeneity, with no detectable evidence of phase separation or elemental segregation at the microscale. Further insight into the local structural order is provided by high-angle annular dark-field scanning transmission electron microscopy (HAADF-STEM). The room temperature HAADF-STEM is done for (Ni$_{0.2}$Mg$_{0.2}$Co$_{0.2}$Cu$_{0.2}$Zn$_{0.2}$)(Cr$_{0.2}$Mn$_{0.2}$Fe$_{0.2}$Ga$_{0.2}$Ti$_{0.2}$)$_2$ O$_4$. The atomic-resolution image acquired along the [103] and [100] zone axis (Fig. 3(a, b)) displays well-resolved atomic columns with characteristic intensity contrast, reflecting the cubic spinel framework. The corresponding fast Fourier transform (FFT) pattern exhibits distinct reflections consistent with cubic symmetry, corroborating the formation of a spinel structure. Simulated HAADF-STEM images, shown alongside the experimental data, reproduce the key features of the observed contrast, further supporting the spinel structural features. The periodic lattice arrangements are in excellent agreement with the expected cation distribution in a spinel lattice, indicating that the extremely high configurational disorder does not disrupt the long-range crystallographic order. Atomic-scale elemental mapping of the same sample is shown in Figure 3(c). It highlights an excellent chemical homogeneity in the nanoscale in a highly disordered sample.

\begin{figure*}
    \centering
    \includegraphics[width=1.00\linewidth]{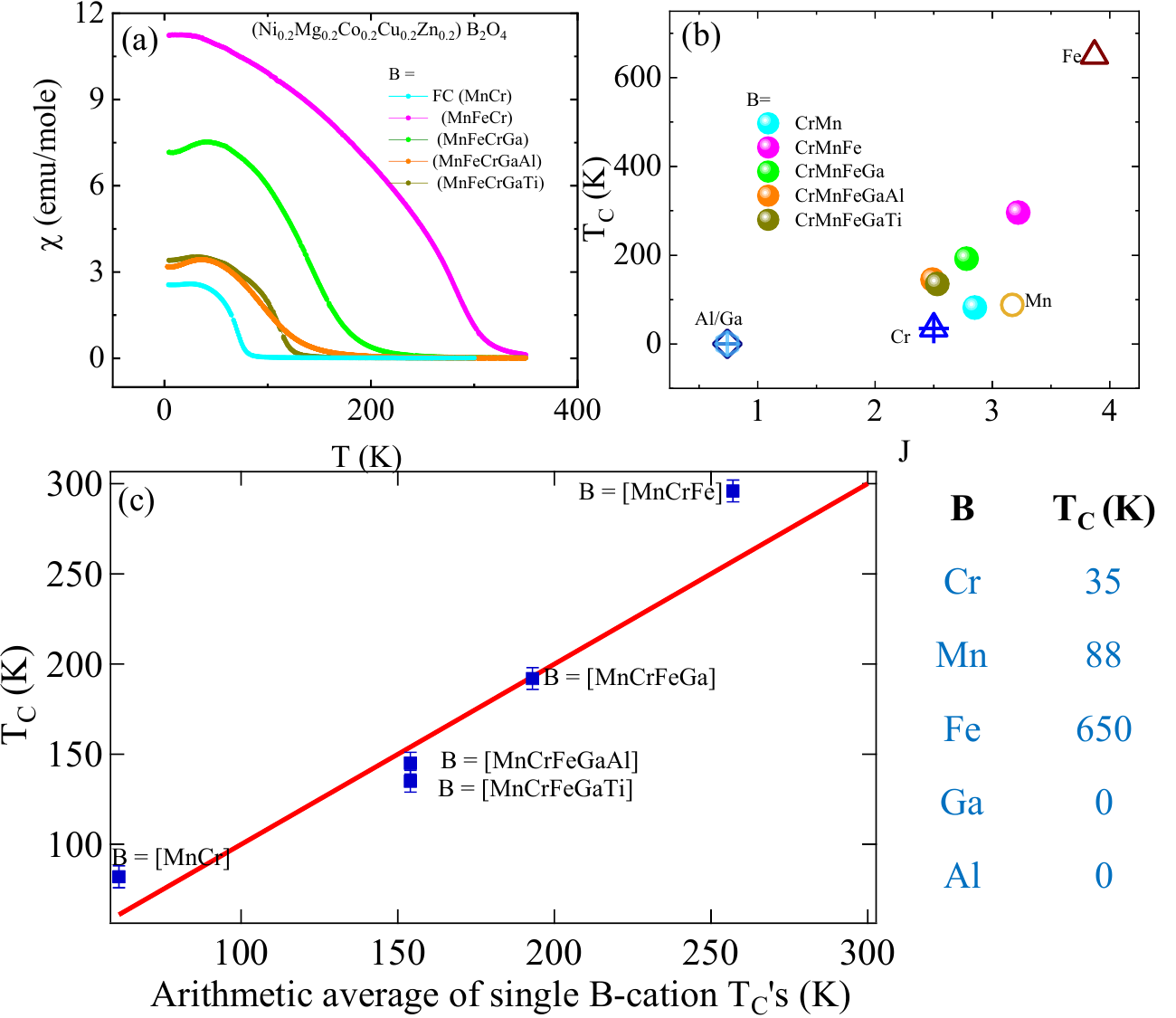}
    \caption{Temperature variation of the FC magnetic susceptibility for (Ni$_{0.2}$Mg$_{0.2}$Co$_{0.2}$Cu$_{0.2}$Zn$_{0.2}$)B$_2$O$_4$ materials, where B = Mn$_{0.5}$Cr$_{0.5}$, Mn$_{0.33}$Cr$_{0.33}$Fe$_{0.33}$, Mn$_{0.25}$Cr$_{0.25}$Fe$_{0.25}$Ga$_{0.25}$, and Mn$_{0.2}$Cr$_{0.2}$Fe$_{0.2}$Ga$_{0.2}$Ti$_{0.2}$. (b) Ferrimagnetic transition temperatures (T$_C$) as a function of total angular momentum (J) for all the samples. (c) It highlights the composition-dependent T$_C$ for all the samples. Upon introducing multicomponent B-site configurations, a linear, nearly arithmetic dependence of the T$_C$ on the magnetic transition temperatures of the corresponding single B-site high-entropy systems is observed with T$_{C}$ (multi-B-site) = $\sum_i x_i T_{Ci,single-B-cation}$. The solid line highlights the theoretically expected T$_C$ using the aforementioned relation. The T$_C$ values for single-cationic B-site compositions are highlighted on the right side of the figure (c). Also, the evolution of $J$ mirrors that of T$_C$, exhibiting a nearly arithmetic compositional dependence consistent with: J = $\sum_i x_i J_i$. Magnetic transition temperatures for single B-site compositions are taken from ref \cite{10, 11}}
    \label{fig5}
\end{figure*}

To further elucidate the local structural environment and its evolution with varying B-site cations in the AB$_2$O$_4$ spinel framework, Raman spectroscopy was employed (Fig. 4). The Raman spectra exhibit four prominent bands centered at 176, 310, 510, and 650 cm$^{-1}$, closely resembling the characteristic vibrational features of conventional spinels such as NiFe$_2$O$_4$ and previously reported high-entropy cubic spinel oxides \cite{10, 11}. According to group theory, spinel structures crystallizing in the Fd-3m space group are expected to display five Raman-active modes, given by:
\begin{equation}
\quad A_{1g} (R) + E_g (R) + 3F_{2g} (R)
\label{eq1}
 \end{equation}
 These modes primarily originate from oxygen sublattice vibrations. In particular, the high-frequency A${1g}$ mode (600–720 cm$^{-1}$) is associated with symmetric stretching of oxygen atoms along the tetrahedral M$^{2+}$–O bonds. The E$g$ mode (250–360 cm$^{-1}$) corresponds to symmetric bending vibrations of oxygen atoms around octahedral M$^{3+}$ cations. The low-frequency F${2g}$(1) mode (160–220 cm$^{-1}$) arises from translational motion of tetrahedral cations with respect to oxygen, while the higher-frequency F${2g}$(2) and F$_{2g}$(3) modes (440–590 cm$^{-1}$) are attributed to asymmetric stretching and bending vibrations of the oxygen framework \cite{34}. The presence of a shoulder near 650 cm$^{-1}$ (associated with the A$_{1g}$ mode) and broadening suggests a distribution of local bonding environments arising from multiple cations occupying both tetrahedral and octahedral sites. Figure 4 (b) shows the evolution of octahedra (B-site) linked E$_g$ modes with B-site variation (average B-site cationic radius) having fixed A-site composition. A shift of the E$_g$ band toward higher wavenumbers is observed. This trend is expected, as the E$_g$ vibrational mode is primarily linked with the B-site cations in the spinel structure; therefore, modifications in the B-site environment lead to a systematic hardening of the E$_g$ mode, resulting in its shift to higher wavenumbers.
 
 Temperature-dependent magnetization measurements were carried out under a field-cooled (FC) protocol for all these samples (Figure 5 (a)). FC-susceptibility data for B = Mn$_{0.33}$Cr$_{0.33}$Fe$_{0.33}$ is reproduced from \cite{16}. For compounds with single B cations: non-magnetic B-site systems (for instance Ga, Al) remain paramagnetic down to low temperatures, whereas magnetic cations (Cr, Mn, Fe) induce clear antiferromagnetic (AFM)/ferrimagnetic transitions with varying T$_C$ \cite{10, 11}. Upon introducing multicomponent B-sites, all mixed-cation compositions retain robust long-range magnetic ordering with well-defined ferrimagnetic transitions. Figure 5 (b) and (c) highlight the T$_C$'s of all the samples.  It reveals a striking compositional averaging behavior in multicomponent B-site spinels. Specifically, the experimentally measured T$_C$ values closely match those estimated from the arithmetic mean of the T$_C$ values of the corresponding single-B-site parent compounds, T$_{C}$ (multi-B-site) = $\sum_i x_i T_{Ci}$(single-B-cation). For example, the measured T$_C$ of (Ni$_{0.2}$Mg$_{0.2}$Co$_{0.2}$Cu$_{0.2}$Zn$_{0.2}$)(MnCrFeGa)$_2$O$_4$ is in close agreement with the averaged value derived from (Ni$_{0.2}$Mg$_{0.2}$Co$_{0.2}$Cu$_{0.2}$Zn$_{0.2}$)Mn$_2$O$_4$, (Ni$_{0.2}$Mg$_{0.2}$Co$_{0.2}$Cu$_{0.2}$Zn$_{0.2}$)Cr$_2$O$_4$,(Ni$_{0.2}$Mg$_{0.2}$Co$_{0.2}$Cu$_{0.2}$
 Zn$_{0.2}$)Fe$_2$O$_4$, and (Ni$_{0.2}$Mg$_{0.2}$Co$_{0.2}$Cu$_{0.2}$Zn$_{0.2}$)Ga$_2$O$_4$. This remarkable trend is consistently observed across all investigated compositions. This behavior is particularly remarkable because the extreme disorder on both sites simultaneously introduces multiple competing exchange pathways, strong local structural distortions, and the absence of a crystallographically unique magnetic sublattice. To understand this behavior, we evaluated the total angular momentum ($J$) using the effective magnetic moments ($\mu_{\mathrm{eff}}$). Notably, the evolution of $J$ mirrors that of T$_C$, exhibiting a nearly arithmetic compositional dependence consistent with: $J$ = $\sum_i x_i J_i$. Within a mean-field approximation, the ordering temperature T$_C$$\propto$ J${_{ex}}$J(J+1), where J${_{ex}}$ is the size of the exchange integral, and J represents total angular momentum. In conventional spinel oxides, magnetism is governed by the exchange interactions between A and B-site cations (J${_{AB}}$ ), with contributions from B-B interactions (J${_{BB}}$). For the high-entropy spinel oxides, we have already observed that magnetism and the order parameter follow the mean-field model \cite{16}. Therefore, in the present high-entropy spinels, the chemically disordered A and B sublattices generate a statistically averaged exchange landscape in which no single magnetic ion or exchange pathway dominates. Consequently, J${_{ex}}$ can be regarded as an effective medium parameter arising from the ensemble average of competing local exchange interactions. Importantly, although the local magnetic environment is highly heterogeneous, the A–B exchange network remains magnetically percolative, thereby sustaining coherent long-range ferrimagnetic order. This averaging mechanism naturally explains the observed linear scaling of both T$_C$ and J, even in the maximally disordered composition, (Ni$_{0.2}$Mg$_{0.2}$Co$_{0.2}$Cu$_{0.2}$Zn$_{0.2}$)(Cr$_{0.2}$Mn$_{0.2}$Fe$_{0.2}$Ga$_{0.2}$Ti$_{0.2}$)$_2$O$_4$. These results reveal a remarkable emergent predictability in ferrimagnetic (colinear magnetic structure) high-entropy spinel oxides: despite extreme configurational disorder and competing microscopic interactions, the bulk magnetic ordering temperature follows a simple arithmetic average of the T$_C$ values. 
\begin{figure*}
  
 \centering
    \includegraphics[width=1.00\linewidth]{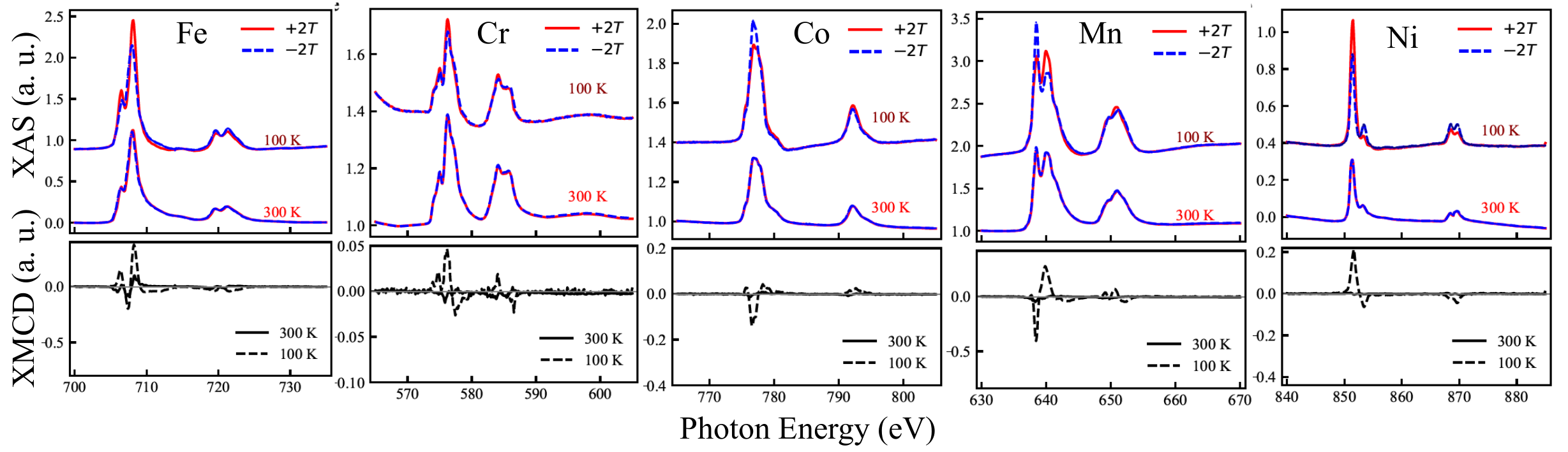}
    \caption{XAS spectra at the L$_{2,3}$ edge of Fe, Cr, Co, Mn, and Ni elements of the (Ni$_{0.2}$Mg$_{0.2}$Co$_{0.2}$Cu$_{0.2}$Zn$_{0.2}$)(Cr$_{0.2}$Mn$_{0.2}$Fe$_{0.2}$Ga$_{0.2}$Ti$_{0.2}$)$_2$ O$_4$ collected at 300 K and 100 K at 2 T with circular polarization at normal incidence. The lower panel shows the XMCD spectra obtained after subtraction of the XAS spectra taken at +2 T and -2 T magnetic fields. }
    \label{figXAS}
\end{figure*}

\begin{figure*}
    \centering
    \includegraphics[width=0.75\linewidth]{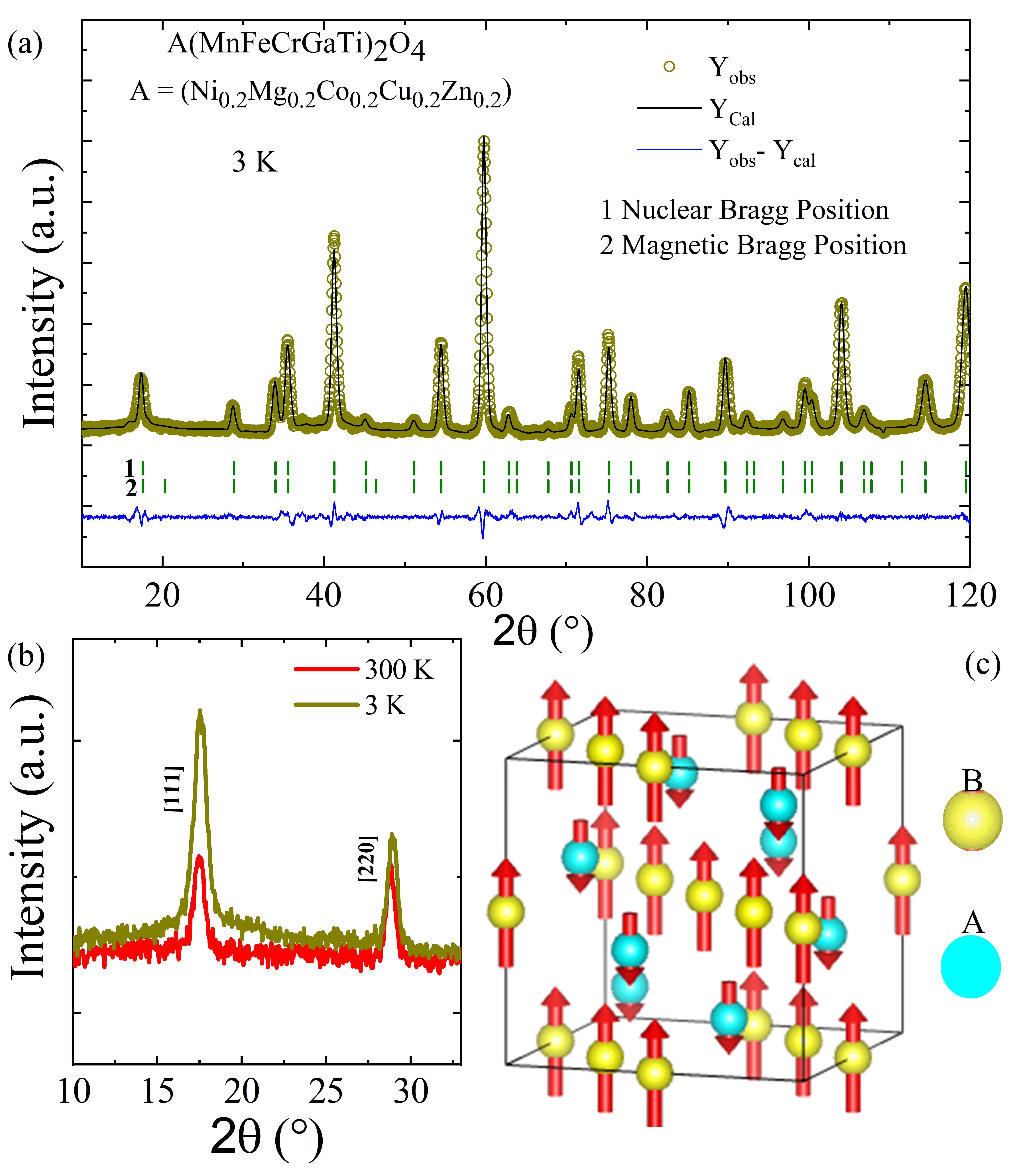}
    \caption{(a) Rietveld refinement of the NPD pattern collected at 3 K for compositionally most complex sample, (Ni$_{0.2}$Mg$_{0.2}$Co$_{0.2}$Cu$_{0.2}$Zn$_{0.2}$)(Cr$_{0.2}$Mn$_{0.2}$Fe$_{0.2}$Ga$_{0.2}$Ti$_{0.2}$)$_2$ O$_4$. (b) Comparison of NPD patterns collected at 300 and 3 K, highlighting the enhanced peak intensities at low temperature, confirming the long-range magnetic ordering. The magnetic structure is indexed with a propagation vector k = (0, 0, 0), indicating a commensurate magnetic ground state. (c) Representation of the magnetic structure. This demonstrates robust ferrimagnetic ordering even in the presence of extreme configurational disorder.}
    \label{fig7}
\end{figure*}

To further probe the local electronic structure, oxidation states, crystal-field environments, and magnetic responses of individual transition-metal cations, we have investigated the soft X-ray absorption spectroscopy (XAS) and X-ray magnetic circular dichroism (XMCD) in the compositionally most complex sample, (Ni$_{0.2}$Mg$_{0.2}$Co$_{0.2}$Cu$_{0.2}$Zn$_{0.2}$)
(Cr$_{0.2}$Mn$_{0.2}$Fe$_{0.2}$Ga$_{0.2}$Ti$_{0.2}$)$_2$ O$_4$. The upper row of Figure \ref{figXAS} shows the absorption spectra taken with circularly polarized light at the L$_{2,3}$ edge of Fe, Cr, Co, Mn, and Ni elements, in a field of +2T and -2T. The measurements were done at two temperatures above (300 K) and below (100 K) the magnetic transition temperature. To assign the site occupancies of the different elements in the sample, the XAS spectral shape obtained at 300 K has been compared with earlier published multiplet calculations performed on similar systems \cite{5}. From the comparative study, Ni${^2+}$ and Cr${^3+}$ are found to occupy the octahedral site, while Fe${^3+}$ spectra resemble the site occupancy for both octahedral and tetrahedral sites, and Co XAS spectra mainly match Co${^2+}$ in tetrahedral geometry. Mn L-edge spectrum (S5 in the SI) is analyzed by comparing the experimental data with simulated spectra for Mn$^{3+}$ in octahedral coordination and Mn$^{2+}$ in tetrahedral sites (using a CTM4XAS pack-
age\cite{35}). By iteratively refining the relative contributions of the two valence states, the experimental spectrum was best reproduced using a weighted combination of approximately 45\% Mn$^{3+}$ (octahedral) and 55\% Mn$^{2+}$ (tetrahedral) \cite{35,36,37} . The Ti $L_{2,3}$ spectrum is characteristic of Ti$^{4+}$ in octahedral environment (S5 in the SI). Our room-temperature X-ray photoelectron spectroscopy (XPS, S6 in the SI) measurements further support the oxidation states inferred from the XAS analysis. Bottom row of Figure \ref{figXAS} shows the difference between the two XAS spectra taken at +2T and -2T field, which corresponds to the XMCD signal. The sign change of the XMCD signal confirms the ferrimagnetic ordering in the sample, a positive XMCD signal for the octahedral geometry, and a negative signal for the tetrahedral geometry, clearly seen for Ni and Co, respectively. The XMCD signal reverses sign, from positive to negative and vice versa, confirming that Fe and Mn occupy both octahedral and tetrahedral sites.

\begin{table}[h!]
\centering
\caption{Magnetic Moments ($\mu_{B}$/atom )calculated by means of the
sum rules from XMCD spectra obtained at T = 100 K for Fe, Ni, Mn, Co, and Cr L-edges. }
\label{Sumrule}
\begin{threeparttable}
\begin{tabular}{c c c c c}
\hline
Element & $m_l$ & $m_s$ & $m_{\mathrm{tot}}$ & $N_h$ \\
\hline

Fe  & -0.034 & 0.652  & 0.618  & 5.0 \\
Ni  & 0.075  & 0.727  & 0.802  & 2.0 \\
Mn  & 0.002  & -0.021 & -0.019 & 5.0 \\
Co  & 0.031  & -0.364 & -0.333 & 4.0 \\
Cr  & 0.124  & 0.396  & 0.520  & 7.0 \\

\hline
\end{tabular}

\begin{tablenotes}
\footnotesize
\item  The error for each magnetic moment is $\pm 10\%$.
\end{tablenotes}

\end{threeparttable}
\end{table}

The values for the effective spin and orbital magnetic moments were obtained from the areas of the L$_{3}$ and L$_{2}$ peaks of the XMCD spectra from the sum rule \cite{39}. 
 \begin{align}
m_l &= -\frac{4}{3}\cdot \frac{q}{r}\cdot \frac{N_h}{\sigma} \\
m_s &= -\frac{6p - 4q}{r}\cdot \frac{N_h}{\sigma}
\end{align}
where N$_{h}$  is the number of holes, p, q, and r are the areas of the XMCD L$_{3}$ and L$_{2}$ regions, and the total area of the isotropic XAS after polynomial background correction. $\sigma$ is the degree of circular polarization, which is beamline-dependent, (0.77). The m$_s$ is the actual spin magnetic moment. In the Table \ref{Sumrule}, the magnetic moment values obtained from the sum-rule analysis for the different elements of the sample are displayed. Data were measured using circularly polarized light and field switching ($\pm 2T$).

\begin{table}[h]
\small
\centering
\caption{Atomic and structural parameters obtained from Rietveld refinement of 3 K neutron powder diffraction (NPD) data for compositionally complex sample, (Ni$_{0.2}$Mg$_{0.2}$Co$_{0.2}$Cu$_{0.2}$Zn$_{0.2}$)(Cr$_{0.2}$Mn$_{0.2}$Fe$_{0.2}$Ga$_{0.2}$Ti$_{0.2}$)$_2$ O$_4$.}
\label{tab:NPD3K}
\begin{tabular}{lll}
\hline
\textbf{Parameter} & \textbf{T = 3 K} & \textbf{} \\
\hline
\textbf{Space Group} & \textit{Fd$\bar{3}$m} & Cubic spinel \\
$a = b = c$ (\AA) & 8.3950 (1) & Lattice parameter \\
\hline
\multicolumn{3}{l}{\textbf{A-site Cations (0.125, 0.125, 0.125)}} \\
Mg Occupancy & 0.22 (2) & B$_\text{iso}$ = 0.67 (6) \\
Cu Occupancy & 0.20 & B$_\text{iso}$ = 0.67 (6)\\
Zn Occupancy & 0.20 & B$_\text{iso}$ = 0.67 (6)\\
Co Occupancy & 0.12 (2)  & B$_\text{iso}$ = 0.67 (6) \\
Mn Occupancy & 0.17 (2) & B$_\text{iso}$ = 0.67 (6) \\
Fe Occupancy & 0.07 (2) & B$_\text{iso}$ = 0.67 (6) \\
\hline
\multicolumn{3}{l}{\textbf{B-site Cations (0.5, 0.5, 0.5)}} \\
Fe Occupancy & 0.31 & B$_\text{iso}$ = 0.16 (3) \\
Cr Occupancy & 0.40 & B$_\text{iso}$ = 0.16 (3) \\
Mn Occupancy & 0.23 (2) & B$_\text{iso}$ = 0.16 (3) \\
Co Occupancy & 0.08 (2) & B$_\text{iso}$ = 0.16 (3) \\
Ni Occupancy & 0.20 & B$_\text{iso}$ = 0.16 (3) \\
Ti Occupancy & 0.40 & B$_\text{iso}$ = 0.16 (3)\\
Ga Occupancy & 0.40 & B$_\text{iso}$ = 0.16 (3) \\
\hline
\multicolumn{3}{l}{\textbf{Oxygen (x, x, x)}} \\
x & 0.2606 (1) & Oxygen positional parameter \\
Occupancy & 1 & Fully occupied \\
B$_\text{iso}$ & 0.45 (3) & -- \\
\hline
B--B distance (\AA) & 2.96807(5) & Octahedral edge \\
A--B distance (\AA) & 3.48037(6) & Tetra--octahedral \\
$\chi^2$ & 5.34 &  \\
$R_p$ & 5.60 &  \\
$R_{wp}$ & 7.46 &  \\
$R_{mag}$ & 3.80 &  \\
M(T$_d$) ($\mu_B$) & 0.78 (4) & A-site magnetic moment \\
M(Oct.) ($\mu_B$) & 1.91 (5) & B-site magnetic moment \\
\hline
\end{tabular}
\end{table}
To understand the nature of the long-range magnetic ordering, we performed the NPD measurements for the most disordered sample (Ni$_{0.2}$Mg$_{0.2}$Co$_{0.2}$Cu$_{0.2}$Zn$_{0.2}$)(Cr$_{0.2}$Mn$_{0.2}$Fe$_{0.2}$Ga$_{0.2}$Ti$_{0.2}$)$_2$O$_4$. The refinement of the RT and 3 K NPD patterns (Figure 7 (a)) confirm a single-phase cubic spinel structure with space group Fd-3m and a lattice parameter of a = 8.3950 (1). Refinement of the oxygen sublattice indicates full occupancy, confirming a stoichiometric oxide without detectable oxygen vacancies. The large contrast in neutron scattering lengths among the constituent elements (e.g., Ni: 1.03 fm, Fe: 0.945 fm, Mn: -0.373 fm, Co: 0.249 fm, O: -0.5803 fm) enables sensitivity to site-specific occupancies despite the high compositional complexity \cite{9}. To constrain the refinement, X-ray absorption spectroscopy (XAS) and X-ray magnetic circular dichroism (XMCD) results were incorporated. Ni is found to occupy the octahedral (16d) sites, whereas Mg, Cu, and Zn reside predominantly on the tetrahedral (8a) sites, while Mn is distributed over both sublattices (Table 1). Also, a small amount of Fe and Co is distributed over both sublattices. The refined cation distribution is consistent with the XAS and XMCD studies.

Strikingly, despite the extreme configurational disorder and the absence of a well-defined magnetic sublattice, the 3 K NPD data reveal clear long-range magnetic ordering (Fig. 7 (b)). All magnetic reflections are indexed with a propagation vector k = (0, 0, 0), indicating a commensurate magnetic structure. The absence of the (200) magnetic Bragg reflection at $2\theta = 20.35^\circ$ rules out non-collinear spin arrangements such as Yafet–Kittel-type canting, while the pronounced (400) reflection at $2\theta = 41.21^\circ$ supports a collinear ferrimagnetic configuration \cite{40}. Within the mean-field framework discussed above, these results provide direct microscopic validation of an effective exchange interaction governing the system. Despite a highly heterogeneous distribution of local superexchange pathways (J$_{AB}$ and J$_{BB}$), the magnetic structure remains globally coherent, consistent with a statistically averaged J${_{ex}}$ that sustains long-range order. Also, recent studies indicate that high-entropy alloys, despite their extreme chemical disorder, exhibit magnetic behavior that is well described within a mean-field framework \cite{16, 41}. The refined magnetic moments at the tetrahedral (M$\mathrm{T_d}$) and octahedral (M$_\mathrm{Oct}$) sites (Table 1) are close to the value obtained from the XMCD data. Together, these observations establish that even in the limit of maximal chemical complexity, the magnetic ground state is governed by an emergent, averaged interaction landscape. This provides a microscopic foundation for the linear tunability of $T_C$ observed across compositions, demonstrating that robust ferrimagnetic order can arise from, rather than be hindered by, extreme configurational disorder.

 \section{Conclusions}
In conclusion, the high-entropy stabilization route emerges here not merely as a strategy for phase formation but as a fundamentally new paradigm for designing collective functionality in complex oxides. By deliberately embedding extreme configurational disorder via multiple principal cations, entropy transforms chemical complexity from a destabilizing perturbation into a thermodynamic driving force that generates emergent magnetic behavior inaccessible in conventional ordered materials. Within the spinel platform, we systematically tuned the B-site configuration of the fixed high-entropy A-site lattice,(Ni$_{0.2}$Mg$_{0.2}$Co$_{0.2}$Cu$_{0.2}$Zn$_{0.2}$)B$_2$O$_4$, to uncover how magnetic order evolves in the presence of competing exchange interactions and highly disordered local environments. A comprehensive combination of XRD, SEM, STEM, XAS, XPS magnetization measurements, XMCD, and neutron powder diffraction unequivocally demonstrates the formation of chemically homogeneous single-phase high-entropy spinels that sustain long-range ferrimagnetic order despite extreme cationic disorder. By creating multicomponent B-site configurations, we demonstrate that the magnetic transition temperature (T$_C$) and total angular momentum (J) of the resulting HEOs follow a nearly linear, arithmetic average of the corresponding single B-cation HEOs, a remarkably simple collective response. Interestingly, this trend persists even in highly complex (Ni$_{0.2}$Mg$_{0.2}$Co$_{0.2}$Cu$_{0.2}$Zn$_{0.2}$)(Cr$_{0.2}$Mn$_{0.2}$Fe$_{0.2}$Ga$_{0.2}$Ti$_{0.2}$)$_2$O$_4$, where no dominant magnetic ion, unique exchange pathway, or chemically ordered magnetic sublattice exists. The bulk magnetic ordering temperature follows a simple effective-medium scaling law, governed by averaged exchange interactions (mean-field model) and angular momentum. Also, detailed low-temperature XMCD and neutron diffraction measurements reveal robust long-range collinear ferrimagnetic ordering. These observations provide microscopic evidence that long-range magnetic correlation can emerge from the statistical averaging of exchange interactions within a magnetically percolating network, rather than requiring chemically ordered superexchange pathways. In this sense, disorder itself becomes the organizing principle governing the collective magnetic ground state. The discovery of a predictable scaling relationship among composition, J values, and magnetic ordering temperature reveals a striking emergent simplicity hidden within extreme chemical complexity. More broadly, this work establishes a compositional route for rationally engineering magnetic transition temperatures in high-entropy spinels, providing a framework for the predictive design of functional magnetic oxides beyond conventional dilute or ordered systems. 

 \section{SUPPORTING INFORMATION}
Experimental methods, figures for x-ray
diffraction, energy dispersive x-ray spectroscopy, magnetization data, X-ray photoelectron spectroscopy data, parameters used in the multiplet theory calculations for the x-ray absorption analysis.

\section{Acknowledgments}
S. M. acknowledges financial support from the ANRF, Government of India, for the Advanced Research Grant (ANRF/ARG/2025/004870/PS). S. M. also acknowledges the financial support from the SPARC, Ministry of Education, Govt. of India (Project no. P4139). D.P. acknowledges the French national Transmission Electron Microscopy and Atomic Probe (METSA) network for the TEM facilities and the PAMEC platform for the technical support. UGC-DAE CSR, Mumbai, is also acknowledged for the Neutron diffraction experiments.

\end{document}